\documentclass[a4paper]{article}

\usepackage{INTERSPEECH2022}
\usepackage{amsthm,amsmath,amssymb}
\usepackage{mathrsfs}
\usepackage{amsmath,graphicx}
\usepackage{multirow}
\usepackage{subfigure}

\title{Tiny-Sepformer: A Tiny Time-Domain Transformer Network for Speech Separation}
\name{Jian Luo$^1$, Jianzong Wang$^{1*}$\thanks{*Corresponding author: Jianzong Wang, jzwang@188.com}, Ning Cheng$^1$, Edward Xiao$^2$, Xulong Zhang$^1$, Jing Xiao$^1$}
\address{$^1$ Ping An Technology (Shenzhen) Co., Ltd. \\$^2$ Aquinas International Academy, CA, USA}
\email{jzwang@188.com}

\begin{document}

\maketitle
\begin{abstract}
Time-domain Transformer neural networks have proven their superiority in speech separation tasks. However, these models usually have a large number of network parameters, thus often encountering the problem of GPU memory explosion. In this paper, we proposed Tiny-Sepformer, a tiny version of Transformer network for speech separation. We present two techniques to reduce the model parameters and memory consumption: (1) Convolution-Attention (CA) block, spliting the vanilla Transformer to two paths, multi-head attention and 1D depthwise separable convolution, (2) parameter sharing, sharing the layer parameters within the CA block. In our experiments, Tiny-Sepformer could greatly reduce the model size, and achieves comparable separation performance with vanilla Sepformer on WSJ0-2/3Mix datasets.
\end{abstract}

\section{Introduction}
\label{sec:intro}

Single-channel multi-speaker speech separation is a significant speech task in real-world applications. Robust speech separation could improve the performance of downstream tasks, such as speaker identification, speech recognition, \textit{etc}~\cite{Zhu2020Identify, Luo2021Unidirectional}. However, speech separation is a difficult task, often known as the cocktail-party problem~\cite{cocktail_ref}. People have made great efforts on deep learning models~\cite{zhang2021singer,Wang2020VoiceFilterLite,zhang2022Singer,Tzinis2020Two,Chen2021Continuous}, which were proposed to advance the progress of this tough task.

Traditional speech separation approaches often transform the mixture signals to time-frequency domain and estimate the clean spectrogram of each speaker from the mixture spectrogram. 
TasNet~\cite{Luo2017TasNet} directly models the audio signal in the time-domain. Conv-TasNet~\cite{Luo2019ConvTasNet} replaces the LSTM of TasNet with 1-D dilated convolutions and it stacks deep convolutional blocks, to model the long-term dependency~\cite{Luo2021Multi,Luo2021Cross}. To promote the efficiency of handling long time-domain sequence, dual-path frameworks were presented. DPRNN~\cite{Luo2020Dual} splits the sequence into small chunks, and applies intra and inter chunk operations iteratively. DPTNet~\cite{Chen2020Dual} introduces Transformer into the recurrent network of DPRNN, and outperforms the vanilla DPRNN. Sepformer~\cite{Subakan2021Attention} was proposed as a RNN-free neural network. The intra and inter chunk operations of masking network are solely based on Transformer, to capture both local and long-term information.

Despite the remarkable achivements of the above Transformer models, they still encounter with some tough problems. One is that they usually have large network parameters, thus often resulting in GPU memory explosion. 
In this work, we focus on the reduction of network parameters and GPU memory consumption.

Recently, many memory or time efficient attention-based models have been proposed. In Linformer~\cite{Wang2020Linformer}, the self-attention mechanism is approximated by a low-rank matrix, resulting a linear complexity Transformer.
Performer~\cite{Choromanski2020Performers} used a FAVOR+ method to model kernelizable attention mechanism efficiently, instead of sparsity or low-rankness. DF-Conformer~\cite{Koizumi2021DFConformer} integrates Conformer~\cite{Gulati2020Conformer} layers with FAVOR+ mechanism into the mask prediction network of Conv-TasNet. Most of these works implemented a efficient system by reducing the self-attention from quadratic to linear complexity, but still have a large amount of model parameters.

In this work, inspired by Lite-Transformer~\cite{Wu2020Lite}, we used Convolution-Attention (CA) block into the masking network, which splits the layer into convolution path and attention path parallelly. The convolution path has much less parameters than the attention path. Moreover, the convolution part of CA is 1D separable convolution~\cite{Hannun2019Sequence}, which could further reduce the computation. Besides the separable convolution, we also applied parameter sharing technique~\cite{Lan2020ALBERT,Chi2021Audio}. All of the layer parameters within one IntraCA/InterCA network are shared, but we do not share the parameters across different IntraCA/InterCA networks. 
In summary, our proposed Tiny-Sepformer has two major contributions to speech separation: (1) CA network, (2) parameter sharing.

\section{Tiny-Sepformer}
\label{sec:tinysep}

In this work, we propose the Tiny-Sepformer, a tiny Transformer network for speech separation. Tiny-Sepformer is a time-domain masking approach, and is composed of three modules: (1) an encoder $\theta_{enc}$, making convolutions on time-domain mixture-signal $X = (x_1,x_2,...,x_T)$ to the feature representation $H = (h_1,h_2,...,h_T)$, where $T$ is the length of time-domian signal, (2) a masking network $\theta_{mask}$, employing dual-path tiny Convolution-Attention (CA) blocks $\theta_{ca}$ on $H$, to estimate $K$ mask matrices $M_{1:k} = \left\{M_1, M_2, ..., M_{K}\right\}$ for each of the $K$ speakers in the mixture-signal, and (3) a decoder $\theta_{dec}$, reconstructing the separated signals $\hat{X}_{1:k}=\left\{\hat{X}_1, \hat{X}_2, ..., \hat{X}_{K}\right\}$ in the time domain by multiplicating the masks $M_K$ with $H$ for each of the $K$ speaker.

\begin{figure*}[htbp]
	\begin{center}
		\centerline{\includegraphics[width=0.8\linewidth]{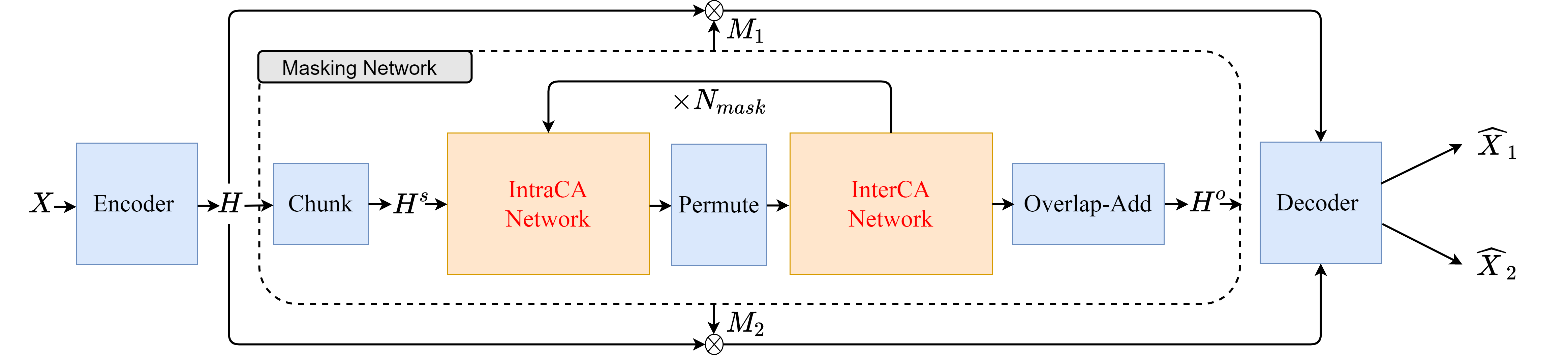}}
		\caption{The Model Architecture of Tiny-Sepformer}
		\label{fig1}
	\end{center}
\end{figure*}

\subsection{Model Architecture}
\label{sec:architecture}
The overall model architecture is depicted in Figure~\ref{fig1}. The time-domain mixture-signal $X$ is firstly fed into the encoder $\theta_{enc}$. The encoder $\theta_{enc}$ is a single 1D-convolutional layer, followed by ReLU activation. The hidden feature representation $H$ is extracted through the encoder.
\begin{equation}
	H = \theta_{enc}(X) = \mathrm{ReLU}(\mathrm{Conv1D}(X))
\end{equation}
After that, the encoder output $H$ is used as the input of the masking network $\theta_{mask}$, to produce mask matrices $M_{1:K}$.
\begin{equation}
	M_{1:K} = \theta_{mask}(H)
\end{equation}
Finally, the mask matrices $M_{1:K}$ and hidden features of encoder $H$ are as the input of the decoder $\theta_{dec}$. The decoder $\theta_{dec}$ is a transposed convolutional layer with the same stride and kernel size of the encoder $\theta_{enc}$. The decoder outputs the separated signal $\hat{X}_{1:K}$ from each source $\left\{1, 2, ..., K\right\}$.
\begin{equation}
	\hat{X}_{k} = \theta_{dec}(M_{k}, H) = \mathrm{Conv1D^T}(M_{k}\odot H)
\end{equation}
In which, $\mathrm{Conv1D^T}$ is the transposed convolution, and $\odot$ is denoted as element-wise multiplication.
The objective of model training is maximizing the Scale-Invariant Source-to-Noise Ratio (SI-SNR). We use Utterance-level Permutation Invariant Training (uPIT)~\cite{Yu2017Permutation,Kolbak2017Multi} loss during training to deal with the label permutation problem.

\subsection{Masking Network}
\label{sec:mask}
The masking network consists of three steps: (1) pre-processing and chunking, (2) the proposed tiny Convolution-Attention (CA) blocks $\theta_{ca}$, and (3) post-processing and overlap-add.

Firstly, the encoder feature sequence $H$ is normalized with layer normalization, and processed by a linear layer to produce the $H^d$ with dimension $T\times D$.
\begin{equation}
	H^d = \mathrm{Linear}(\mathrm{LayerNorm}(H))
\end{equation}
After pre-processing, the hidden feature sequence $H^d$ is processed by the segmentation operation. The segmentation splits two dimensional $H^d\in T\times D$ into three dimensional chunks $H^s\in T_{S}\times S\times D$ with $50\%$ overlap. In which, $S$ is the chunk size of segmentation, and $T_{S}$ is the chunk number of segmentation result. The chunked $H^s$ is fed into several CA blocks $\theta_{ca}$, which will be detailed in Section~\ref{sec:ca}. The CA blocks will be performed by $N_{mask}$ times iteratively to produce $H^{ca}$.
\begin{equation}
	H^{ca} = \theta_{ca}(H^s)_{\times N_{mask}}
\end{equation}
The output of CA block $H^{ca}$ remains the same dimension $T_{S}\times S\times D$ with the input $H^s$. Then, $H^{ca}$ is processed by a linear layer with dimension $(D\times K)$ and with using the PReLU activation.
\begin{equation}
	H^{dk} = \mathrm{PReLU}(\mathrm{Linear}(H^{ca}))
\end{equation}
This post-processing operation generates feature maps $H^{dk}\in T_{S}\times S\times (D\times K)$ for each of the $K$ speakers. Afterwards, the Overlap-Add~\cite{Luo2020Dual} is operated on $H^{dk}$, merging the three dimensional chunked sequence into two dimensinal feature sequence $H^o\in T \times (D\times K)$.
\begin{equation}
	M_{1:K} = \mathrm{ReLU}(\mathrm{Linear}(H^o)_{\times 2})
\end{equation}
At the end, this representation $H^{o}$ is through two linear layers and a ReLU activation to obtain the mask matrices $M_{1:K}$ for each of the $K$ speakers. The dimension of each mask $M_K$ is $T\times D$. 

\subsection{Convolution-Attention (CA) Block}
\label{sec:ca}

The Convolution-Attention (CA) Block $\theta_{ca}$ is the core module of masking network. There are two kinds of network in the masking network: (1) IntraCA network $\theta_{ca}^{intra}$, operating within each chunk to model local features, (2) InterCA network $\theta_{ca}^{inter}$, processing between all the chunks to capture long dependency. The $\theta_{ca}$ block are designed to process $\theta_{ca}^{intra}$ network firstly, then permute the feature dimension from $T_{S}\times S\times D$ to $S\times T_{S}\times D$, and process $\theta_{ca}^{inter}$ network at last.
\begin{equation}
\label{eq:ca}
\begin{aligned}
H^{ca} &= \theta_{ca}(H^s) \\
&= \theta_{ca}^{inter}(\mathcal{P}(\theta_{ca}^{intra}(H^s)_{\times N_{intra}}))_{\times N_{inter}}
\end{aligned}
\end{equation}
In which, $\mathcal{P}$ is denoted as the permutation operation. IntraCA $\theta_{ca}^{intra}$ is performed $N_{intra}$ times, and InterCA $\theta_{ca}^{inter}$ will perform by $N_{inter}$ times iteratively.

\begin{figure}[ht]
	\begin{center}
		\centerline{\includegraphics[width=0.75\columnwidth]{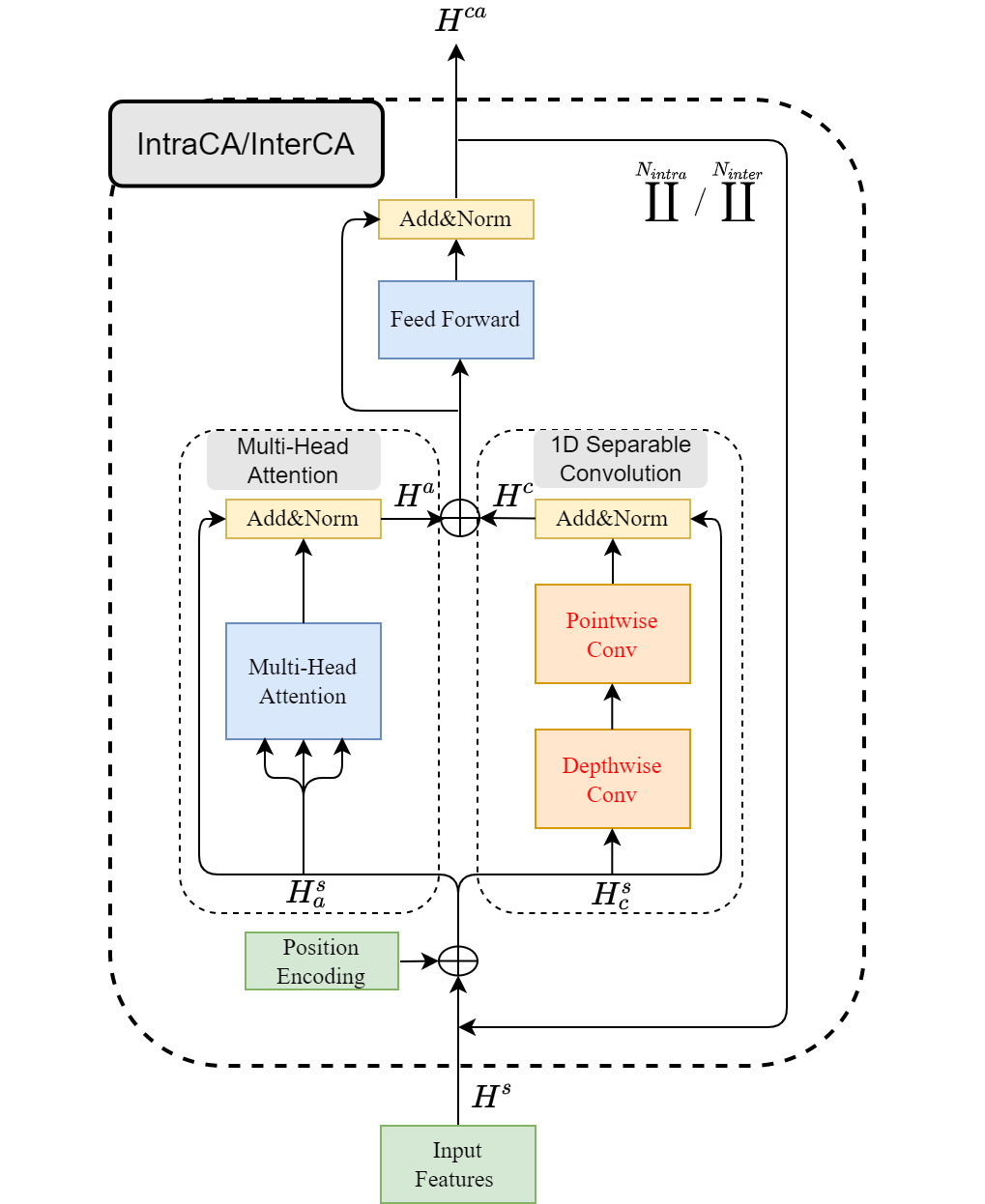}}
		\caption{The Structure of IntraCA/InterCA Network}
		\label{fig2}
	\end{center}
\end{figure}

As depicted in Figure~\ref{fig2}, IntraCA or InterCA network contains two parts: (1) multi-head attention, (2) 1D depthwise separable convolution. IntraCA $\theta_{ca}^{intra}$ and InterCA $\theta_{ca}^{inter}$ have the same network structure, but have different network parameters. Taking IntraCA $\theta_{ca}^{intra}$ as the example, the input feature $H^s\in T_{S}\times S\times D$ is split into two paths: (1) attention feature $H^s_a\in T_{S}\times S\times D_a$, and (2) convolution feature $H^s_c\in T_{S}\times S\times D_c$, where $D_c + D_a = D$. The choices of $D_c$ and $D_a$ will be analyzed in Section~\ref{sec:division}. The attention feature $H^s_a$ is fed into the standard multi-head attention mechanism, followed by layer normalization.
\begin{equation}
	H^a = \mathrm{LayerNorm}(\mathrm{MultiHeadAttention}(H^s_a) + H^s_a)
\end{equation}
Mean while, the other branch $H^s_c$ is processed by the 1D depthwise separable convolution as follows:
\begin{equation}
	H^s_d = \mathrm{DepthwiseConv1D}(H^s_c)
\end{equation}
\begin{equation}
	H^c_p = \mathrm{PointwiseConv1D}(H^s_d)
\end{equation}
followed by residual addition and layer normalization.
\begin{equation}
	H^c = \mathrm{LayerNorm}(H^c_p + H^s_c)
\end{equation}
The 1D depthwise separable convolution could dramaticly reduce the number of network parameters. In addition, the dimension reduction of multi-head attention $D_a$ could also make a light attention layer. 

After that, the output of two path $H^a$ and $H^c$ are concatenated together, feeding into feed-forward layers as follows:
\begin{equation}
	H^{ca}_f = \mathrm{FeedForward}(H^a\oplus H^c)
\end{equation}
In which, $\oplus$ is denoted as concatenated operation. The dimension of feed-forward layer is denoted as $D_f$. Following the feed-forward layer are residual addition and layer normalization, similar to the vanilla Transformer network.
\begin{equation}
	H^{ca} = \mathrm{LayerNorm}(H^{ca}_f + H^a\oplus H^c)
\end{equation}

\subsection{Parameter Sharing}
\label{sec:paramshare}
Another idea of parameters reduction is cross-layer parameter sharing. There are several methods of sharing parameters, such as sharing the feed-forward network parameters, or only sharing attention and convolution parameters. In this work, we propose to share all the layer parameters within IntraCA/InterCA network, but different IntraCA/InterCA networks have different parameters. Therefore, with parameter sharing, the Equation~\ref{eq:ca} is modified as following:
\begin{equation}
H^{ca} = \theta_{ca}(H^s) = \coprod^{N_{inter}}\theta_{ca}^{inter}(\mathcal{P}(\coprod^{N_{intra}}\theta_{ca}^{intra}(H^s)))
\end{equation}
In which, $\coprod^{N}$ means that performing the network layer by $N$ times iteratively, but each iteration has the same layer parameters. With this approach, the parameters of $\theta_{ca}^{intra}$ could decrease to ${1/N_{intra}}$ of the original network, as well as the ${1/N_{inter}}$ parameters for $\theta_{ca}^{inter}$. We share all the layers in IntraCA $\theta_{ca}^{intra}$ and InterCA $\theta_{ca}^{inter}$ respectively, resulting in much fewer network parameters of the Tiny-Sepformer model.

\subsection{Compared with Conformer}
\label{sec:conformer}
The most relevant structure to CA block is Conformer~\cite{Gulati2020Conformer}, which connects the attention and convolution in serial. The motivation of our work is to reduce the model parameters. Therefore, we design to connect the multi-head attention and 1D depthwise separable convolution in parallel. For simplicity, in this section, we denoted the input feature of the block as $H^s\in T\times D$, where $T$ is the number of time steps and $D$ is the channel dimension. Then, the parameters of attention and convolution paths are as Table~\ref{tab1}:

\begin{table}[ht]
	\centering
	\caption{The Model Parameters of Attention and Convolution}
	\scalebox{0.9} {
		\begin{tabular}{c|c}
			\hline
			Network & Parameters \\
			\hline
			Multi-Head Attention & $4\times D^2$\\
			1D Depthwise Separable Convolution & $K\times D+D^2$\\
			Attention+Convolution in Serial (Conformer) & $K\times D+5\times D^2$\\
			\textbf{Attention+Convolution in Parallel (CA)} & $\frac{K}{2}\times D+\frac{5}{4}\times D^2$\\
			\hline
		\end{tabular}
	}
	\label{tab1}
\end{table}

The query, key, value and multi-head projection matrix has $D^2$ parameters respectively. Thus, for multi-head attention, the parameters are $4\times D^2$. The separable convolution has a depthwise convolutional layer with kernel size $K$ on each channel individually, and a pointwise convolutional layer on each frame independently but across all channels $D$. Therefore, for separable convolution, the parameters are $K\times D+D^2$. If connecting Attention+Convolution in Serial (Conformer) and keeping the feature dimension as $D$, their parameters are added together ($K\times D+5\times D^2$). For our CA block (Attention+Convolution in Parallel), the input feature is split into two branches: $D_a+D_c=D$. In Table~\ref{tab1}, we set $D_a=D_c=\frac{1}{2}\times D$ for simplicity, then the parameters are reduced to $\frac{K}{2}\times D+\frac{5}{4}\times D^2$. Through the channel division operation, the parameters of CA block are much less than vanilla Transformer and Conformer, and still keep the feature dimension as $D$ into next feedforward layer. 

\section{Experiment}
\label{sec:exp}

\subsection{Dataset}
\label{sec:data}
We use the WSJ0-2mix and WSJ0-3mix~\cite{Hershey2016Deep} datasets to evaluate our Tiny-Sepformer model. These two datasets were generated by randomly selecting utterances from the WSJ0 corpus, and mixing them with two and three speakers. $30$ hours of training, $10$ hours of validation and $5$ hours of test speech dataset were used for all the experiments. All the speech were downsampled to $8$kHz in the data pre-processing. 

\begin{table*}[ht]
	\centering
	\caption{Different Layer Configurations of Tiny-Sepformer, Results on WSJ0-2Mix and WSJ0-3Mix}
	\scalebox{0.92} {
		\begin{tabular}{p{3.0cm}|p{1.0cm}|p{1.0cm}|p{1.0cm}|p{1.5cm}|p{1.2cm}|p{1.5cm}|p{1.2cm}|p{1.2cm}|p{1.0cm}}
			\hline\hline
			\multirow{2}{*}{\textbf{Model}} & \multirow{2}{*}{$N_{mask}$} & \multirow{2}{*}{$N_{intra}$} & \multirow{2}{*}{$N_{inter}$} & \multicolumn{2}{c|}{\textbf{WSJ0-2Mix}} & \multicolumn{2}{c|}{\textbf{WSJ0-3Mix}} & \multirow{2}{*}{\textbf{Sharing}} & \multirow{2}{*}{\textbf{Param}} \\
			\cline{5-8}
			& & & & SI-SNRi & SDRi & SI-SNRi & SDRi & & \\
			\hline
			Sepformer-16~\cite{Subakan2021Attention} & 2 & 4 & 4 & 14.08 & 15.01 & 12.29 & 13.19 & No & 13.0M \\
			Sepformer-32~\cite{Subakan2021Attention} & 2 & 8 & 8 & 15.08 & 16.04 & 12.67 & 13.71 & No & 25.7M \\
			Sepformer-32~\cite{Subakan2021Attention} & 4 & 4 & 4 & 15.07 & 16.02 & 13.02 & 14.01 & No & 25.7M \\
			\hline
			Tiny-Sepformer-16 & 2 & 4 & 4 & 14.29 & 15.13 & 12.87 & 13.85 & No & 10.2M \\
			Tiny-Sepformer-32 & 2 & 8 & 8 & 15.09 & 16.03 & 14.38 & 15.36 & No & 20.0M \\
			Tiny-Sepformer-32 & 4 & 4 & 4 & 15.10 & \textbf{16.07} & \textbf{14.50} & \textbf{15.53} & No & 20.0M \\
			\hline
			Tiny-SepformerS-16 & 2 & 4 & 4 & 13.51 & 14.22 & 12.38 & 13.21 & Yes & \textbf{2.9M} \\
			Tiny-SepformerS-32 & 2 & 8 & 8 & 14.66 & 15.39 & 12.77 & 13.63 & Yes & \textbf{2.9M} \\
			Tiny-SepformerS-32 & 4 & 4 & 4 & \textbf{15.16} & 15.98 & 13.91 & 14.79 & Yes & 5.3M \\
			\hline\hline
		\end{tabular}
	}
	\label{tab2}
\end{table*}


\begin{table}[ht]
	\centering
	\caption{Different Channel Divisions of Tiny-Sepformer-32 and Tiny-SepformerS-32, Results on WSJ0-2Mix}
	\scalebox{0.9} {
		\begin{tabular}{p{2.7cm}|p{1.4cm}|p{1.4cm}|p{1.1cm}|p{0.7cm}}
			\hline\hline
			\multirow{2}{*}{\textbf{Model}} & \textbf{IntraCA} & \textbf{InterCA} & \multicolumn{2}{c}{\textbf{WSJ0-2Mix}} \\
			\cline{4-5}
			& $D_{c}$, $D_{a}$ & $D_{c}$, $D_{a}$ & SI-SNRi & SDRi \\
			\hline
			Tiny-Sepformer-32 & 128, 128 & 128, 128 & 15.10 & 16.07 \\
			Tiny-Sepformer-32 & \textbf{192}, 64 & 64, \textbf{192} & \textbf{15.46} & \textbf{16.36} \\
			Tiny-Sepformer-32 & 64, \textbf{192} & \textbf{192}, 64 & 14.97 & 15.87 \\
			\hline
			Tiny-SepformerS-32 & 128, 128 & 128, 128 & 15.16 & 15.98 \\
			Tiny-SepformerS-32 & \textbf{192}, 64 & 64, \textbf{192} & 15.21 & 16.06 \\
			Tiny-SepformerS-32 & 64, \textbf{192} & \textbf{192}, 64 & 15.13 & 15.95 \\
			\hline\hline
		\end{tabular}
	}
	\label{tab3}
\end{table}

\subsection{Model Configuration}
\label{sec:param}
We conducted all the experiments using the SpeechBrain toolkit~\cite{Mirco2021SpeechBrain}. For encoder $\theta_{enc}$, the 1D-convolutional layer has $256$ filters, a kernel size of $16$, and a stride factor of $8$. For masking network $\theta_{mask}$, the dimension $D$ of pre-processing is $256$. The segmentation splits the chunks with chunk size $S = 250$ with $50\%$ overlap. For decoder $\theta_{dec}$, the transposed convolutional layer has the same kernel size and stride factor with the encoder.

We use dynamic mixing (DM)~\cite{Zeghidour2021Wavesplit} and speed perturbation ($95\%$-$105\%$ randomly) for data augmentation. Adam optimizer~\cite{Kingma2014Adam} was used with learning rate of $1.5e^{-4}$. We also use automatic mixed precision~\cite{Micikevicius2018Mixed} to speed up training.

\subsection{Result}
\label{sec:result}
In our experiments, Scale-Invariant Source-to-Noise Ratio improvement (SI-SNRi) and Signal-to-Distortion Ratio improvement (SDRi)~\cite{Vincent2006Performance} are used as the evaluating metrics. As listed in Table~\ref{tab2}, we firstly explored different configuration of CA blocks number $N_{mask}$, IntraCA layers number $N_{intra}$, and InterCA layers number $N_{inter}$. The results indicated that our Tiny-Sepformer models achieve comparable separation performance with vanilla Sepformer on both WSJ0-2mix and WSJ0-3mix datasets, but have fewer model parameters. Furthermore, using the method of parameter sharing, the Tiny-SepformerS model could greatly reduce the model size, but with only a little performance degradation. All of the models are trained for $150$ epochs with batchsize $1$ on one NVIDIA V100 GPU card with $16$ GB memory. In particular, the $32$-layers Sepformer-32 is trained within $16$ GB GPU, instead of $32$ GB. We set a limit of training signal length $T$ to $64K$, to control the GPU memory consumption. For fair comparison, our Tiny-Sepformer-32 and Tiny-SepformerS-32 models used the same setting of this length limit.

\subsection{Channel Division}
\label{sec:division}
As shown in Table~\ref{tab3}, we also investigate different channel divisions of multi-head attention $D_a$ and separable convolution $D_c$ in IntraCA and InterCA network respectively. The best Tiny-Sepformer-32 and Tiny-SepformerS-32 models ($N_{mask}=N_{intra}=N_{inter}=4$) in Table~\ref{tab2} are used. The feed-forward dimension $D_f$ is $1024$, and the number of attention heads is $8$. The convolutional kernel size of IntraCA is $51$, and the kernel size of InterCA is $11$. The results demonstrated that large $D_c$ dimension ($D_c=192$) in IntraCA and $D_a$ dimension ($D_a=192$) in InterCA are better. The convolution helps the model to capture local information within each chunk, and the attention on the contrary models global context among all the chunks.

\subsection{Attention Weights Analysis}
\label{sec:analysis}

\begin{figure}[ht]
	\centering
	\subfigure[Attention Weights of IntraCA] { \label{fig3a}
		\includegraphics[width=0.47\columnwidth]{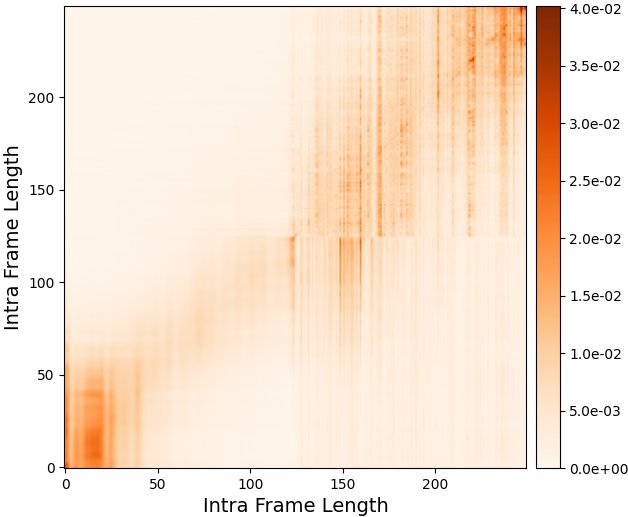}
	}
	\subfigure[Attention Weights of InterCA] { \label{fig3b}
		\includegraphics[width=0.47\columnwidth]{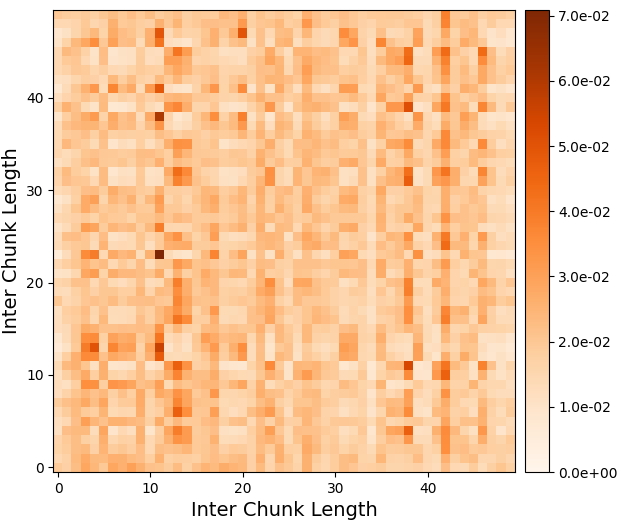}
	}
	\caption{Attention Weights Analysis of Tiny-Sepformer-32 \newline (IntraCA $D_{a}=64$, InterCA $D_{a}=192$)}
	\label{fig3}
\end{figure}

To further make the analysis of the function of convolution and attention paths, we plotted the weights of attention matrix in IntraCA and InterCA respectively. Best configuration in Table~\ref{tab3} (IntraCA $D_{a}=64$, InterCA $D_{a}=192$) of Tiny-Sepformer-32 is used in Figure~\ref{fig3}. We found that the attentions of IntraCA are likely to gather together on the diagonal line (in Figure~\ref{fig3a}). It means that large parts of attention could be replaced with local convolution in IntraCA. On the contrary, the attentions of InterCA are distributed globally among all the chunks (in Figure~\ref{fig3b}). The huge difference between these two attention maps indicated that it is reasonable to assign more channel dimensions to $D_c$ for IntraCA and $D_a$ for InterCA.

\section{Conclusion}
\label{sec:conclusions}
In this work, we propose Tiny-Sepformer, a tiny Transformer network for speech separation. The Convolution-Attention network splits the features to two paths, and replaces one path with light 1D separable convolution. We also shared the layer parameters within the CA block, to further reduce the model parameters. The proposed Tiny-Sepformer achieves comparable separation results, and has relatively small model size. In addition, we found that large convolution channels within the chunk and more attention channels among the chunks could further improve the performance. The analysis of attention matrix weights explain the reason for the choice of this channel division configuration.
For future works, we are also interested in exploring better model structure for fast inference speed. Memory consumption and time cost are both crucial for real-world application, like streaming speech separation scenarios.

\section{Acknowledgement}
This paper is supported by the Key Research and Development Program of Guangdong Province under grant No.2021B0101400003. Corresponding author is Jianzong Wang from Ping An Technology (Shenzhen) Co., Ltd (jzwang@188.com).

\bibliographystyle{IEEEtran}
\bibliography{refs}

\end{document}